\documentclass[showpacs,preprintnumbers,amsmath,amssymb]{revtex4}

\usepackage{graphicx}
\usepackage{dcolumn}
\usepackage{bm}
\usepackage{color}

\newcommand{\bq}{\begin{equation}}
\newcommand{\eq}{\end{equation}}
\newcommand{\bqa}{\begin{eqnarray}}
\newcommand{\eqa}{\end{eqnarray}}
\newcommand{\nn}{\nonumber \\}

\def\be     {\begin{equation}}
\def\ee     {\end{equation}}
\def\bea        {\begin{eqnarray}}
\def\eea        {\end{eqnarray}}
\def\bnn    {\begin{eqnarray*}}
\def\enn    {\end{eqnarray*}}

\usepackage[
pagebackref=false,
colorlinks=true,
linkcolor=blue,
urlcolor=blue,
filecolor=black,
citecolor=red,
pdfstartview=FitV,
pdftitle={},
pdfauthor={},
pdfsubject={},
pdfkeywords={},
pdfpagemode=None,
bookmarksopen=true
]{hyperref}

\begin{document}

%
%\title{Introduction of UV QFT information to IR emergent dual gravity away from
%  quantum criticality
%}
%

\title{
  Emergent dual holographic description as
  a non-perturbative generalization of
  the
  %perturbative
  Wilsonian renormalization group
  %Emergent dual holographic description as generalization of the perturbative
  %Wilsonian renormalization-group method into a non-perturbative way
}

\author{Ki-Seok Kim$^{a,c}$, Shinsei Ryu$^{b}$, and Kanghoon Lee$^{c,a}$}
\affiliation{$^{a}$Department of Physics, POSTECH, Pohang, Gyeongbuk 37673, Korea \\ $^{b}$Department of Physics, Princeton University, Princeton, New Jersey, 08540, USA \\ $^{c}$Asia Pacific Center for Theoretical Physics (APCTP), Pohang, Gyeongbuk 37673, Korea}

\email[Ki-Seok Kim: ]{tkfkd@postech.ac.kr}
\email[Shinsei Ryu: ]{shinseir@princeton.edu}
\email[Kanghoon Lee: ]{kanghoon.lee1@gmail.com}

\date{\today}

\begin{abstract}
%Since holographic duality had been conjectured, it has been immediately
%suggested that dynamics along the emergent extra-dimensional space describes a
%renormalization group (RG) flow of the corresponding quantum field theory (QFT).
In holographic duality, 
dynamics along the emergent extra-dimensional space describes
a renormalization group (RG) flow of the corresponding quantum field theory (QFT).
Following this idea, we develop an emergent holographic description of a QFT,
where not only the information of the RG flow is introduced into an IR
holographic dual effective field theory (HDEFT),
but also the UV information of the QFT
is encoded in the HDEFT through the IR boundary condition.
In particular, we argue that this dual holographic construction is self-consistent
within the assumption of bulk locality,
showing the following two aspects:
The solution of the Hamilton-Jacobi equation is given by the IR boundary
effective action,
and the Ward identity involving the QFT energy-momentum tensor current is satisfied naturally.
We discuss the role of the RG $\beta$-function in the bulk effective dynamics of the metric tensor near a conformally invariant fixed point.
%
%As a result, we find that the universal lower bound of the ratio between the shear viscosity and the entropy can be reduced, where the RG $\beta-$function plays a similar role of higher-derivative curvature terms in Einstein gravity.
%
We claim that this emergent dual gravity theory generalizes the perturbative Wilsonian RG framework into a non-perturbative way.
%
%the $AdS_{D+1}$$-$$CFT_{D}$ duality conjecture away from quantum criticality.
%
%{\color{red}(Non-conformal holographic dual is known previously, so we need to be careful here.)}
%
\end{abstract}

%\pacs{71.10.Hf, 71.30.+h, 71.10.-w, 71.10.Fd}

\maketitle

%
%{\color{red} (The title may be ok, but is there a better title?)}
%

\section{Introduction}

Non-perturbative approaches to quantum field theory (QFT)
are highly sought after to study strongly-coupled problems.
%It is highly desirable to develop a non-perturbative theoretical framework to study strongly-coupled quantum field theory (QFT).
Specifically,
%Here,
%being non-perturbative
%means that
we need a framework in which
relevant quantum corrections to
self-energies and vertices are self-consistently re-summed
in the infinite order
in the renormalization group (RG) sense.
%even if this non-perturbative re-summation is not exact but approximate.
%This is so even if this non-perturbative re-summation is not exact but approximate.
The ${\it AdS}_{D+1}$/${\it CFT}_{D}$
duality conjecture
\cite{Holographic_Duality_I,Holographic_Duality_II,Holographic_Duality_III,Holographic_Duality_IV}
has been regarded as a non-perturbative theoretical framework,
where a non-perturbative RG flow of
%an effective
a UV field theory is realized through the evolution
along the extra-dimensional space \cite{Holographic_Duality_V,Holographic_Duality_VI,Holographic_Duality_VII}. Here, $D$ is the spacetime dimension.

The holographic approach has been applied to various strongly-coupled
problems,
such as
confinement and chiral symmetry breaking in quantum chromodynamics
\cite{Holographic_Duality_II,Holographic_Duality_III}
and superconductivity and non-Fermi liquids in condensed matter physics
\cite{Nakayama:2013is,Superconductivity_Holography,Benincasa:2012wu,Erdmenger:2013dpa,OBannon:2015cqy,Erdmenger:2016jjg,NFL_Holography,
  FQHE_Holography,MIT_Holography,Graphene_AdS_Description},
and provided remarkable solutions,
%to these strongly-coupled problems,
e.g., emergent physics of effective hydrodynamics
\cite{Holographic_Liquid_Son_I,Holographic_Liquid_Son_II,Holographic_Liquid_Son_III,Holographic_Liquid_Son_IV}.
These results, in view of universality,
can in principle be applicable to a wide class of problems.
%This non-perturbative theoretical framework gives remarkable solutions
%sometimes,
%for example, emergent physics of
%effective hydrodynamics
%\cite{Holographic_Liquid_Son_I,Holographic_Liquid_Son_II,Holographic_Liquid_Son_III,Holographic_Liquid_Son_IV}.
However,
there still exists an unsatisfactory point:
it is not entirely clear how to relate UV microscopic degrees of
freedom
with IR emergent macroscopic observables.
%{\color{red}may elaborate on this point..}
To overcome this difficulty,
various approaches have been tried
%{\color{red}make english smooth here..}
to derive an effective holographic dual field theory based on RG transformations \cite{RG_Holography_I,RG_Holography_II,RG_Holography_III,RG_Holography_IV,RG_Holography_V,RG_Holography_VI,RG_Holography_VII,RG_Holography_VII_Q,
RG_Holography_VIII,RG_Holography_IX,RG_Holography_X,RG_Holography_XI,RG_Holography_XII,RG_Holography_XIII,RG_Holography_XIV,RG_Holography_XV,
RG_Holography_XVI,RG_Holography_XVII,RG_Holography_XVIII,RG_Holography_XIX,RG_Holography_XX,SungSik_Holography_I,SungSik_Holography_II,
SungSik_Holography_III,Vasudev_Shyam_I,Vasudev_Shyam_II,Vasudev_Shyam_III,Vasudev_Shyam_IV,Einstein_Klein_Gordon_RG_Kim,Einstein_Dirac_RG_Kim,
RG_GR_Geometry_I_Kim,RG_GR_Geometry_II_Kim,Kondo_Holography_Kim,Kitaev_Entanglement_Entropy_Kim,RG_Holography_First_Kim,
Path_Integral_Optimization_I,Path_Integral_Optimization_II,Path_Integral_Optimization_III}.

In this study, we continue to follow these lines of
approach
and develop further
an emergent holographic description of a QFT.
%a quantum field theory (QFT).
%In this study, we develop an emergent holographic description of a quantum field theory (QFT).
In this framework,
not only the information of the RG flow is included into an IR
holographic dual effective field theory,
but also the UV information of the QFT is encoded
%into the holographic dual effective field theory
through the IR boundary condition.
In particular, we argue that,
within the assumption of bulk locality,
this dual holographic construction is self-consistent,
showing the following two aspects:
The solution of the Hamilton-Jacobi equation is given by the IR boundary
effective action, and the Ward identity involving
the QFT energy-momentum tensor current is satisfied naturally.
We discuss the role of the RG $\beta$-function in the bulk effective dynamics of the metric tensor near a conformally invariant fixed point.

Recently, it has been clarified that the Wess-Zumino consistency condition for
the local RG flow of a QFT can be translated into the Hamilton-Jacobi
formulation of a holographic dual effective field theory
\cite{Vasudev_Shyam_I,Vasudev_Shyam_II,Vasudev_Shyam_III,Vasudev_Shyam_IV}.
%{\color{red}(read these papers -- are they related to the paper by Schamik?)}
The present study takes into account
%
%{\color{red}(Is follow the right verb here?)}
%
this internal consistency for the emergent dual holographic description.
The resulting holographic dual effective field theory generalizes
the previous construction,
%
%{\color{red}
%  (Maybe it's better to be a bit more specific about
%  what the previous construction is.)}
%
where an IR boundary condition is introduced as the solution of the Hamilton-Jacobi equation.
This IR boundary condition
%manifests
makes it manifest
how to encode the UV information of the QFT into the IR holographic dual
effective field theory even away from quantum criticality.
This framework thus extends
the ${\it AdS}_{D+1}/{\it CFT}_{D}$ duality conjecture
to systems away from criticality.
More generally, we claim that this emergent dual gravity theory
is a non-perturbative
generalization
of
the perturbative Wilsonian RG framework.
%into a non-perturbative framework.

\section{Emergent dual holography as a renormalization group flow}

%\subsection{Construction of holographic dual effective field theory away from quantum criticality}

\subsection{Construction of holographic dual effective field theory}

%Let us start from the following Euclidean path integral:
%A partition function is given by
The starting point of our analysis is the following Euclidean path integral
in $D$ spacetime dimensions:
\begin{align}
  \label{starting path int}
  Z
  &= \int D \psi_{\alpha}(x) D g_{B}^{\mu\nu}(x)
    ~ \delta\Big(g_{B}^{\mu\nu}(x) - \delta^{\mu\nu}\Big) ~
    \nonumber \\
  &\quad
    \times
    \exp\Big[ - \int d^{D} x \sqrt{g_{B}(x)}
    \Big\{ \mathcal{L}[\psi_{\alpha}(x), g_{B}^{\mu\nu}(x)]
    + \frac{\lambda}{2 N} T^{\mu\nu}(x) \mathcal{G}_{\mu\nu\rho\gamma}^{B}(x) T^{\rho\gamma}(x) \Big\} \Big] .
\end{align}
Here, $\psi_{\alpha}(x)$
($\alpha=1,\cdots, N$)
is a matter field
with its dynamics described by
%the dynamics of which is described by UV effective Lagrangian $\mathcal{L}[\psi_{\alpha}(x), g_{B}^{\mu\nu}(x)]$.
the Lagrangian $\mathcal{L}[\psi_{\alpha}(x), g_{B}^{\mu\nu}(x)]$.
%In this study we focus on relativistic invariance only.
In this study, we focus only on relativistic invariant theories.
$g_{B}^{\mu\nu}(x)$
is a formally introduced background metric,
and enforced to be the flat metric,
$g_{B}^{\mu\nu}(x)=\delta^{\mu\nu}$,
by the delta functional.
The Lagrangian
$\mathcal{L}[\psi_{\alpha}(x), g_{B}^{\mu\nu}(x)]$
is deformed by an effective interaction
that is quadratic in the
energy-momentum tensor
(the $T\bar{T}$ deformation in $D$ spacetime dimensions)
\cite{TTbar_Deformation},
%we consider a deformed theory given by effective interactions between energy-momentum tensors \cite{TTbar_Deformation},
where
$T^{\mu\nu}(x) = \frac{2}{\sqrt{g_{B}(x)}} \frac{\delta }{\delta
  g_{\mu\nu}^{B}(x)} \Big( \sqrt{g_{B}(x)} \mathcal{L}[\psi_{\alpha}(x),
g_{B}^{\mu\nu}(x)] \Big)$,
%is {\color{red}the} energy-momentum tensor.
$\lambda \geq 0$ is the coupling constant,
and
$\mathcal{G}_{\mu\nu\rho\gamma}^{B}(x) \equiv \frac{1}{2}g_{\mu\rho}^{B}(x)
g_{\nu\gamma}^{B}(x) + \frac{1}{2}g_{\nu\rho}^{B}(x) g_{\mu\gamma}^{B}(x) -
\frac{1}{D-1} g_{\mu\nu}^{B}(x) g_{\rho\gamma}^{B}(x)$
is the DeWitt supermetric \cite{DeWitt_Metric}, taking into account transverseness.
%
%{\color{red}(Can we explain the notation? E.g., $(.|.|.)$ represents symmetrization over indices, etc. $\rightarrow$ Kanghoon, please check this out. Frankly speaking, I do not know what this notation means, too.\textcolor{blue}{$\to$ I've replaced the expression to avoid the notation (18).})}
%

The type of theory in \eqref{starting path int}
can be studied by using
the ``recursive RG transformations'';
Performing the functional RG transformation
(with the Hubbard-Stratonovich transformation)
in a recursive way, one can construct an
IR holographic dual effective field theory,
which describes the evolution of the metric tensor
in the RG (energy) scale.
%as emergent gravity
The resulting theory takes the form of emergent gravity,
with emergent extra dimension representing
the RG scale of the problem
\cite{SungSik_Holography_I,SungSik_Holography_II,SungSik_Holography_III,Einstein_Klein_Gordon_RG_Kim,
  Einstein_Dirac_RG_Kim,RG_GR_Geometry_I_Kim,RG_GR_Geometry_II_Kim}.
In the following, we coordinatize the extra dimension by $z$,
and $z=0$ and $z_f$ by convention represent the UV and IR energy scales,
respectively.

The details of the steps
to derive the holographic theory
starting from specific UV quantum field theories
with double trace interactions
can be found in
\cite{SungSik_Holography_I,SungSik_Holography_II,SungSik_Holography_III,Einstein_Klein_Gordon_RG_Kim,
  Einstein_Dirac_RG_Kim,RG_GR_Geometry_I_Kim,RG_GR_Geometry_II_Kim}.
In general,
the functional RG transformations give rise to nonlocal
effective interactions.
Such emergent nonlocal interactions can however
be ``localized''
at the cost of
introducing higher-spin fields to decompose them in a local fashion based on the corresponding group structure \cite{Higher_Spin_Gauge_Theory_I,Higher_Spin_Gauge_Theory_II,Higher_Spin_Gauge_Theory_III,Higher_Spin_Gauge_Theory_IV,Higher_Spin_Gauge_Theory_V}.
%\textcolor{red}{(Sounds plausible, but is there a concrete analysis on this?? $\rightarrow$ I do not know. What I know is the recent work of Ref. \cite{Higher_Spin_Gauge_Theory_V}. There, I am not sure whether they follow this route or not.)}
In other words, integrating over such higher-spin fields
gives rise to
an effective gravity theory
including only up to spin two fields,
but in the presence of
effective nonlocal interactions between gravitons.
In most cases, we will work with
a proper local truncation of
these RG-generated nonlocal terms \cite{Comment_Higher_Spin_GT},
keeping the original form of the effective Lagrangian
as in the conventional RG transformation
\cite{Einstein_Klein_Gordon_RG_Kim,Einstein_Dirac_RG_Kim}.

%We would like to emphasize that this UV-IR mapping is not exact but approximate.
%{\color{red}Mention what part is approximate.}
%In particular, the functional RG transformation gives rise to nonlocal effective interactions generically. Such emergent nonlocal interactions can be ``localized", introducing higher-spin fields to decompose them in a local fashion based on the corresponding group structure \cite{Higher_Spin_Gauge_Theory_I,Higher_Spin_Gauge_Theory_II,Higher_Spin_Gauge_Theory_III,Higher_Spin_Gauge_Theory_IV,Higher_Spin_Gauge_Theory_V}. In other words, integrating over such higher-spin fields, we obtain an effective gravity theory up to spin two but in the presence of effective nonlocal interactions between gravitons. Here, these RG-generated nonlocal terms are truncated in a local way \cite{Comment_Higher_Spin_GT}, keeping the original form of the effective Lagrangian as the conventional RG transformation \cite{Einstein_Klein_Gordon_RG_Kim,Einstein_Dirac_RG_Kim}.

With the locality assumption in mind,
in this paper, we propose
%(Is ``propose'' a precise way to describe the nature of this work?
%$\rightarrow$ I think `derive' sounds somewhat strong since both the
%Einstein-Hilbert action and the RG-$\beta$ function for the metric tensor are
%all just formal. I am not so sure whether they can be well regularized and
%renormalized, where this calculation is essentially based on quantum gravity in
%the `Wilsonian scheme' whatever it means. `construct' would be also O.K. in my
%opinion.)
a generic dual holographic effective theory resulting from the recursive RG
transformations.
It is given by
%The resulting holographic dual effective field theory is given by
\begin{align}
  \label{hol eff th}
  Z
  &= \int D \psi_{\alpha}(x) D g_{\mu\nu}(x,z) D \pi^{\mu\nu}(x,z) D \mathcal{N}(x,z) D \mathcal{N}_{\mu}(x,z) ~ \delta\Big(g^{\mu\nu}(x,0) - g^{\mu\nu}_{B}(x)\Big)
    \nonumber \\
  & \quad \times
    \exp\Big[ - \int d^{D} x \sqrt{g(x,z_{f})} ~ \mathcal{L}[\psi_{\alpha}(x), g^{\mu\nu}(x,z_{f})]
    \nonumber \\
  &\quad\qquad
    - N \int_{0}^{z_{f}} d z \int d^{D} x \Big\{ \pi^{\mu\nu}(x,z) \partial_{z} g_{\mu\nu}(x,z) + \mathcal{N}(x,z) \mathcal{H} + \mathcal{N}_{\mu}(x,z) \mathcal{H}^{\mu} \Big\} \Big] .
\end{align}
Here, the emergent bulk dynamical metric tensor is given by
\bqa && d s^{2} = \Big( \mathcal{N}^{2}(x,z) + \mathcal{N}_{\mu}(x,z) \mathcal{N}^{\mu}(x,z) \Big) d z^{2} + 2 \mathcal{N}_{\mu}(x,z) d x^{\mu} d z + g_{\mu\nu}(x,z) d x^{\mu} d x^{\nu} . \eqa
$\mathcal{N}(x,z)$ and $\mathcal{N}^{\mu}(x,z)$ are the lapse function and the
shift vector, respectively, and $g_{\mu\nu}(x,z)$ is the $D$-dimensional metric
tensor in the Arnowitt-Deser-Misner (ADM) decomposition
\cite{ADM_Hamiltonian_Formulation}.
The dynamics of the metric tensor is governed by
the effective Hamiltonian
\bqa && \mathcal{H} = \frac{\lambda}{2} \frac{1}{\sqrt{g(x,z)}} \pi^{\mu\nu}(x,z) \mathcal{G}_{\mu\nu\rho\gamma}(x,z) \pi^{\rho\gamma}(x,z) + \mathcal{H}_{\beta} + \mathcal{H}_{g} , \eqa
that can be regarded as a generator of the RG transformation
along the $z$ direction.
Here, $\pi^{\mu\nu}(x,z)$ is the momentum that is
canonically conjugate to the metric tensor $g_{\mu\nu}(x,z)$,
and $\mathcal{G}_{\mu\nu\rho\gamma}(x,z)$ is the bulk supermetric tensor.
The first term in this bulk effective Hamiltonian results from the energy-momentum tensor deformation at UV \cite{TTbar_Deformation}.

The last part of this effective Hamiltonian originates from quantum fluctuations of matter fields in the RG transformation, expressed by the vacuum-energy functional of the renormalized effective Lagrangian at a given RG scale $z$,
\bqa \int d^{D} x ~ \mathcal{H}_{g} &=& (-1)^{F} \ln \int D \psi_{\alpha}^{h}(x) \exp\Big\{- \int d^{D} x \sqrt{g(x,z)} \mathcal{L}[\psi_{\alpha}^{h}(x), g^{\mu\nu}(x,z)] \Big\} \nn &\equiv& (-1)^{F} \mbox{tr} \ln \Big\langle
\frac{\partial^{2}}{\partial \psi_{\alpha}(x) \partial \psi_{\beta}(x')} \Big(
\sqrt{g(x,z)} \mathcal{L}[\psi_{\alpha}(x), g^{\mu\nu}(x,z)] \Big) \Big\rangle .
\eqa
Here, $\int D \psi_{\alpha}^{h}(x)$ represents to take high-energy quantum fluctuations of matter fields
%
%where $\langle  \cdots \rangle$ is the expectation value in terms of the effective Lagrangian
%
at a given RG scale $z$,
%{\color{red}($<..>$ is w.r.t. which action? $\rightarrow$ As you know, this is nothing but the inverse of the Green's function. What I am concerned is that the notation itself looks to hold in the non-interacting case if I delete the expectation symbol. There must be a background potential to decompose an effective interaction, for example, $\varphi(x,z)$ in the $\phi^{4}$ theory. Precisely, it corresponds to $(-1)^{F} \mbox{tr} \ln \Big\langle
%\frac{\partial^{2}}{\partial \psi_{\alpha}(x) \partial \psi_{\beta}(x')} \Big(
%\sqrt{g(x,z)} \mathcal{L}[\psi_{\alpha}(x), g^{\mu\nu}(x,z)] \Big) \Big\rangle = (-1)^{F} \ln \int D \psi_{\alpha}^{h}(x) \exp\Big\{- \int d^{D} x \sqrt{g(x,z)} \mathcal{L}[\psi_{\alpha}^{h}(x), g^{\mu\nu}(x,z)] \Big\}$, where the superscript $h$ means `heavy' or 'high-energy'. In the former expression, the superscript `h` is omitted. In summary, the symbol to express the expectation value disappears safely, where the background potential field to decompose an effective interaction is introduced. Let me consider a better expression more.)}
and $F = 0$ ($F = 1$) when the matter fields are bosonic (fermionic).
Performing the
gradient expansion for the metric tensor, one finds the Einstein-Hilbert action
$\mathcal{H}_{g} = \frac{\sqrt{g(x,z)}}{2 \kappa} \Big( R(x,z) - 2 \Lambda
\Big)$, referred to as induced gravity
\cite{Gradient_Expansion_Gravity_I,Gradient_Expansion_Gravity_II}, where higher
curvature terms are not taken into account \cite{Higher_Curvature_GR}.
Here, both the cosmological constant $\Lambda$ and the effective gravitational one $\kappa$
can in principle be determined
by performing the gradient expansion on a general curved spacetime manifold explicitly,
while it can be demanding in practice due to renormalization effects.
%It is demanding to perform the gradient expansion explicitly and to determine the cosmological constant with renormalization effects.
In this study we regard them as input parameters.

The second part of this effective Hamiltonian is given by \bqa &&
\mathcal{H}_{\beta} = - \pi^{\mu\nu}(x,z) \beta_{\mu\nu}^{g}[g_{\mu\nu}(x,z)],
\eqa
where $\beta_{\mu\nu}^{g}[g_{\mu\nu}(x,z)]$ is the RG $\beta$-function
% for the RG flow
of the metric tensor and given by
\bqa && \beta_{\mu\nu}^{g}[g_{\mu\nu}(x,z)] = - \frac{\mathcal{C}_{g}}{N} \mathcal{G}_{\mu\nu\rho\gamma}(x,z) \Big\langle T^{\rho\gamma}(x,z) \Big\rangle . \label{beta_ft} \eqa
Here, $\mathcal{C}_{g}$ is
%the order of $1$ numerical constant,
a numerical constant of order one
and
\bqa && \Big\langle T^{\rho\gamma}(x,z) \Big\rangle = \frac{2}{\sqrt{g(x,z)}} \frac{\delta }{\delta g_{\rho\gamma}(x,z)} \Big\langle \sqrt{g(x,z)} \mathcal{L}[\psi_{\alpha}(x), g^{\mu\nu}(x,z)] \Big\rangle \label{Energy_Momentum_Tensor} \eqa
is the energy-momentum tensor defined
in terms of the effective Lagrangian for the
matter field at a given RG-transformation slice $z$.

Finally,
the last bulk term of the
holographic dual effective field theory
(\ref{hol eff th})
is given by
\begin{align}
\mathcal{H}^{\mu} = 2 \mathcal{D}_{\nu} \pi^{\mu\nu}(x,z),
\end{align}
where $\mathcal{D}_{\nu}$ is the covariant derivative in the ADM decomposition.
This is the generator for diffeomorphism of the $D$-dimensional spacetime.
Performing the path integral for $\mathcal{N}_{\mu}(x,z)$,
we obtain
the constraint
$\mathcal{D}_{\nu} \pi^{\mu\nu}(x,z) = 0$ \cite{Holographic_Ward_Identity}.
%{\color{red}
%
%In the boundary theory,
%
This corresponds to the Ward identity involved with the $D$-dimensional QFT
energy-momentum tensor current at a given $z$.
We will show
that the canonical momentum tensor $\pi^{\mu\nu}(x,z_{f})$ is given by the energy-momentum tensor of the renormalized IR QFT at the IR boundary $z = z_{f}$.
%{\color{red} (Is this sentence correct? $\rightarrow$ Frankly speaking, I do not catch your point. Anyway, to avoid any possible confusion, let me reexpress this sentence as the above.)}

Once again,
the holographic dual effective field theory
\eqref{hol eff th}
can in principle be derived, starting from a given UV field theory,
by following the
recursive RG
procedures 
%
%\textcolor{blue}{(KH: Could you make a comment briefly what is these
%  procedures for beginners?) }
%\textcolor{red}{(SR: it is the recursive RG procedure,
%  which is briefly discussed in the next section. I reworded the sentence
%  slightly -- hope it helps.)}
%
in
\cite{SungSik_Holography_I,SungSik_Holography_II,SungSik_Holography_III,Einstein_Klein_Gordon_RG_Kim,
  Einstein_Dirac_RG_Kim,RG_GR_Geometry_I_Kim,RG_GR_Geometry_II_Kim}.
We expect to end up with
the holographic dual effective field theory
\eqref{hol eff th}.
Instead of pursuing the top-down approach,
we will verify,
in the next section
\ref{Verification for the construction of the holographic dual effective field theory},
that the effective holographic theory \eqref{hol eff th},
once discretized, leads to the recursive RG transformations.
In Section
\ref{Self-consistency of the holographic dual effective field theory in the
  Hamilton-Jacobi formulation},
we further discuss
self-consistency of the holographic dual effective theory,
in particular, the compatibility of the
Callan-Symanzik equation $d\ln Z/dz_f =0$
and the Hamilton (Hamilton-Jacobi) equation of motion
derived from
\eqref{hol eff th}.

Before proceeding to these discussions,
we make a few brief comments here.

First,
we note that if we start from the UV boundary theory
which is conformal as in ${\it AdS}_{D+1}$/${\it CFT}_{D}$,
the beta function vanishes and
we do not have the $\mathcal{H}_{\beta}$
term that is linear in $\pi^{\mu\nu}(x,z)$.
The holographic dual effective theory \eqref{hol eff th}
is more generic, and incorporate the effect of
the non-zero beta function.
%{\color{red}
%Appearance of the linear term on $\pi^{\mu\nu}(x,z)$ is beyond the
%$AdS_{D+1}$$-$$CFT_{D}$ duality conjecture,
%}

Second,
compared with the holographic effective theories
considered in
\cite{SungSik_Holography_I,SungSik_Holography_II,SungSik_Holography_III,Einstein_Klein_Gordon_RG_Kim,
  Einstein_Dirac_RG_Kim,RG_GR_Geometry_I_Kim,RG_GR_Geometry_II_Kim},
Eq.\ \eqref{hol eff th} is written in a covariant way
by incorporating the lapse function and the shift vector.
Taking the limit of $\lambda \rightarrow 0$ with gauge fixing $\mathcal{N}(x,z)
= 1$ and $\mathcal{N}_{\mu}(x,z) = 0$,
we obtain the RG flow of the metric tensor, $\partial_{z} g_{\mu\nu}(x,z) =
\beta_{\mu\nu}^{g}[g_{\mu\nu}(x,z)]$
after the path integral over $\pi^{\mu\nu}(x,z)$.
Solving the RG equation for the metric
with a suitable boundary condition, we obtain
the renormalized metric $g_{\mu\nu}(x,z_f)$ in IR,
which in turn enters in the IR effective
Lagrangian $\mathcal{L}[\psi_{\alpha}(x), g^{\mu\nu}(x,z_{f})]$
and determines the IR boundary condition.
$g_{\mu\nu}(x,z)$ thus needs to be determined self-consistently
%which serves as the IR boundary condition for self-consistency
\cite{Einstein_Klein_Gordon_RG_Kim,Einstein_Dirac_RG_Kim,RG_GR_Geometry_I_Kim,RG_GR_Geometry_II_Kim}.
Both the RG $\beta$-function and the IR boundary condition complete the UV-IR mapping manifestly, which will be more clarified below.

%{\color{red}(I needed to introduce what $N$ is. Please check.)}
%
%{\color{red}(Explain the motivation for $\lambda$...
%  In we don't have $\lambda$, we cannot get something like (30).
%  )}
%
%{\color{red}Is here $S_{UV}$ and UV bdry condition?}

%\subsection{Verification for the construction of the holographic dual effective field theory}
\subsection{The holographic dual effective field theory
and recursive RG transformations}
\label{Verification for the construction of the holographic dual effective field theory}

To verify the above construction
and make a contact with the recursive RG transformations,
we perform the path integral with respect to the canonical momentum $\pi^{\mu\nu}(x,z)$ and obtain the Lagrangian formulation as follows
\begin{align}
  & Z = \int D \psi_{\alpha}(x) D g_{\mu\nu}(x,z) ~ \delta\Big(g^{\mu\nu}(x,0) - \delta^{\mu\nu}\Big) ~ \exp\Bigg[ - \int d^{D} x \sqrt{g(x,z_{f})} ~ \mathcal{L}[\psi_{\alpha}(x), g^{\mu\nu}(x,z_{f})]
    \nn
  & - N \int_{0}^{z_{f}} d z \Bigg\{ - \int d^{D} x \frac{\sqrt{g(x,z)}}{2 \lambda} \mathcal{G}^{\mu\nu\rho\gamma}(x,z) \Bigg( \partial_{z} g_{\mu\nu}(x,z) + \frac{2 \mathcal{C}_{g}}{\sqrt{g(x,z)}} \mathcal{G}_{\mu\nu\alpha\beta}(x,z) \frac{\delta \Big\langle \sqrt{g(x,z)} \mathcal{L}[\psi_{\alpha}(x), g^{\mu\nu}(x,z)] \Big\rangle}{\delta g_{\alpha\beta}(x,z)} \Bigg)
    \nn
  & \qquad \qquad \qquad \qquad \qquad
    \times \Bigg( \partial_{z} g_{\rho\gamma}(x,z) + \frac{2 \mathcal{C}_{g}}{\sqrt{g(x,z)}} \mathcal{G}_{\rho\gamma\alpha'\beta'}(x,z) \frac{\delta \Big\langle \sqrt{g(x,z)} \mathcal{L}[\psi_{\alpha}(x), g^{\mu\nu}(x,z)] \Big\rangle}{\delta g_{\alpha'\beta'}(x,z)} \Bigg)
    \nn
  &
    \qquad \qquad\qquad
    + (-1)^{F} \mbox{tr} \ln \Big\langle \frac{\partial^{2}}{\partial \psi_{\alpha}(x) \partial \psi_{\beta}(x')} \Big( \sqrt{g(x,z)} \mathcal{L}[\psi_{\alpha}(x), g^{\mu\nu}(x,z)] \Big) \Big\rangle \Bigg\} \Bigg] . \label{Recursive_RG_EFT_Continuum}
\end{align}
Here, the normal coordinate system of $d s^{2}(x,z) = d z^{2} + g_{\mu\nu}(x,z)
d x^{\mu} d x^{\nu}$ has been considered with gauge fixing
%given by
$\mathcal{N}(x,z) = 1$ and $\mathcal{N}_{\mu}(x,z) = 0$.
We emphasize that all the essential information of the bulk effective action is
given by the effective renormalized UV field theory
$\mathcal{L}[\psi_{\alpha}(x), g^{\mu\nu}(x,z)]$ at a given RG scale $z$ in a
self-consistent way
as it should. %be.

Now, we make the extra-dimensional space $z$ discrete
and introduce the discrete coordinate $k$
that represents the RG transformation step,
\bqa && Z = \int D \psi_{\alpha}(x) D g_{\mu\nu}^{(k)}(x) ~
\delta\Big(g^{\mu\nu}_{(0)}(x) - \delta^{\mu\nu}\Big) ~ \exp\Bigg[ - \int d^{D}
x \sqrt{g_{(f)}(x)} ~ \mathcal{L}[\psi_{\alpha}(x), g^{\mu\nu}_{(f)}(x)] \nn &&
- N (d z) \sum_{k = 1}^{f} \Bigg\{ - \int d^{D} x \frac{\sqrt{g_{(k-1)}(x)}}{2
  \lambda} \mathcal{G}^{\mu\nu\rho\gamma}_{(k-1)}(x) \Bigg( g_{\mu\nu}^{(k)}(x)
- g_{\mu\nu}^{(k-1)}(x) \nn && + \frac{2 \mathcal{C}_{g}}{\sqrt{g_{(k-1)}(x)}}
\mathcal{G}_{\mu\nu\alpha\beta}^{(k-1)}(x) \frac{\delta }{\delta
  g_{\alpha\beta}^{(k-1)}(x)} \Big\langle \sqrt{g_{(k-1)}(x)}
\mathcal{L}[\psi_{\alpha}(x), g^{\mu\nu}_{(k-1)}(x)] \Big\rangle \Bigg) \Bigg(
g_{\rho\gamma}^{(k)}(x) - g_{\rho\gamma}^{(k-1)}(x) \nn && + \frac{2
  \mathcal{C}_{g}}{\sqrt{g_{(k-1)}(x)}}
\mathcal{G}_{\rho\gamma\alpha'\beta'}^{(k-1)}(x) \frac{\delta }{\delta
  g_{\alpha'\beta'}^{(k-1)}(x)} \Big\langle \sqrt{g_{(k-1)}(x)}
\mathcal{L}[\psi_{\alpha}(x), g^{\mu\nu}_{(k-1)}(x)] \Big\rangle \Bigg) \nn && +
(-1)^{F} \mbox{tr} \ln \Big\langle \frac{\partial^{2}}{\partial \psi_{\alpha}(x)
  \partial \psi_{\beta}(x')} \Big( \sqrt{g_{(k-1)}(x)}
\mathcal{L}[\psi_{\alpha}(x), g^{\mu\nu}_{(k-1)}(x)] \Big) \Big\rangle \Bigg\}
\Bigg] . \label{Recursive_RG_EFT} \eqa
Here, $d z$ is an energy scale for the RG transformation, and $z_{f} = f d z$ is
the energy scale where quantum fluctuations of matter fields are integrated out.
%
%Here, we point out that there is an issue, when one transforms derivative expressions from ``continuous'' to ``discrete'' or vice versa, whether to follow the ``Ito'' or ``Stratonovich'' scheme \cite{Ito_Stratonovich}. In the current context, we are following the Ito scheme for continuous-discrete transformation, where the Jacobian factor does not arise in the continuum approximation.
%
The above discrete expression is consistent with the recursive RG-transformation method of our previous studies \cite{Einstein_Klein_Gordon_RG_Kim,Einstein_Dirac_RG_Kim,RG_GR_Geometry_I_Kim,RG_GR_Geometry_II_Kim}.

Focusing on the first iteration of the RG transformation, we consider $k = 1$ and obtain
\bqa && Z = \int D \psi_{\alpha}(x) D g_{\mu\nu}^{(0)}(x) D g_{\mu\nu}^{(1)}(x)
~ \delta\Big(g^{\mu\nu}_{(0)}(x) - \delta^{\mu\nu}\Big) ~ \exp\Bigg[ - \int
d^{D} x \sqrt{g_{(1)}(x)} ~ \mathcal{L}[\psi_{\alpha}(x), g^{\mu\nu}_{(1)}(x)]
\nn && - N (d z) \Bigg\{ - \int d^{D} x \frac{\sqrt{g_{(0)}(x)}}{2 \lambda}
\mathcal{G}^{\mu\nu\rho\gamma}_{(0)}(x) \Bigg( g_{\mu\nu}^{(1)}(x) -
g_{\mu\nu}^{(0)}(x) \nn && + \frac{2 \mathcal{C}_{g}}{\sqrt{g_{(0)}(x)}}
\mathcal{G}_{\mu\nu\alpha\beta}^{(0)}(x) \frac{\delta }{\delta
  g_{\alpha\beta}^{(0)}(x)} \Big\langle \sqrt{g_{(0)}(x)}
\mathcal{L}[\psi_{\alpha}(x), g^{\mu\nu}_{(0)}(x)] \Big\rangle \Bigg) \Bigg(
g_{\rho\gamma}^{(1)}(x) - g_{\rho\gamma}^{(0)}(x) \nn && + \frac{2
  \mathcal{C}_{g}}{\sqrt{g_{(0)}(x)}}
\mathcal{G}_{\rho\gamma\alpha'\beta'}^{(0)}(x) \frac{\delta }{\delta
  g_{\alpha'\beta'}^{(0)}(x)} \Big\langle \sqrt{g_{(0)}(x)}
\mathcal{L}[\psi_{\alpha}(x), g^{\mu\nu}_{(0)}(x)] \Big\rangle \Bigg) \nn && +
(-1)^{F} \mbox{tr} \ln \Big\langle \frac{\partial^{2}}{\partial \psi_{\alpha}(x)
  \partial \psi_{\beta}(x')} \Big( \sqrt{g_{(0)}(x)}
\mathcal{L}[\psi_{\alpha}(x), g^{\mu\nu}_{(0)}(x)] \Big) \Big\rangle \Bigg\}
\Bigg] . \eqa Taking the $\lambda \rightarrow 0$ limit,
we find the RG flow of
the metric tensor is given by $g_{\mu\nu}^{(1)}(x) = g_{\mu\nu}^{(0)}(x) - \frac{2
  \mathcal{C}_{g}}{\sqrt{g_{(0)}(x)}} \mathcal{G}_{\mu\nu\alpha\beta}^{(0)}(x)
\frac{\delta }{\delta g_{\alpha\beta}^{(0)}(x)} \Big\langle \sqrt{g_{(0)}(x)}
\mathcal{L}[\psi_{\alpha}(x), g^{\mu\nu}_{(0)}(x)] \Big\rangle$.
This
turns out to be identical with the RG transformation
at the one-loop level
\cite{Einstein_Klein_Gordon_RG_Kim,Einstein_Dirac_RG_Kim,RG_GR_Geometry_I_Kim,RG_GR_Geometry_II_Kim}. The next iteration step is to separate slow and fast degrees of freedom for both $g_{\mu\nu}^{(1)}(x)$ and $\psi_{\alpha}(x)$, and to perform the path integral with respect to all fast degrees of freedom. As a result, $g_{\mu\nu}^{(1)}(x)$ renormalizes into $g_{\mu\nu}^{(2)}(x)$, where the RG $\beta-$function is now given by $g_{\mu\nu}^{(1)}(x)$. The renormalized metric tensor $g_{\mu\nu}^{(2)}(x)$ appears into the renormalized effective Lagrangian. Repeating this RG-transformation procedure, we obtain Eq. (\ref{Recursive_RG_EFT}) in a discrete form and Eq. (\ref{Recursive_RG_EFT_Continuum}) in a continuum expression.

We recall that this RG-transformation procedure is analogous to the numerical renormalization-group (NRG) method \cite{NRG_Wilson,NRG_Review}.
First, we perform exact diagonalization in the so-called Wilson chain, and
truncate the resulting Hilbert space into its low-energy subspace, regarded to
be coarse graining.
Then, we increase the system size, adding one site into the Wilson chain,
and repeat the RG procedure until it converges.
The present recursive RG-transformation method implements the NRG philosophy in
an analytic way,
representing renormalization effects of coupling functions as an RG flow of the
metric tensor through the emergent extra-dimensional space with a single-trace
deformation of the energy-momentum tensor current.
This above demonstrates
how the present holographic dual effective field theory
takes into account quantum corrections in a non-perturbative way,
i.e., all-loop order resummed through the RG flow in the extra-dimensional space \cite{Kondo_Holography_Kim}.

\subsection{Self-consistency of the holographic dual effective field theory in the Hamilton-Jacobi formulation}
\label{Self-consistency of the holographic dual effective field theory in the Hamilton-Jacobi formulation}

We now study
the self-consistency of the
holographic dual effective field theory,
%including from quantum criticality.
including the case of non-conformal theories,
$\beta^g_{\mu\nu}\neq 0$.
%To argue self-consistency of this holographic dual effective field theory away
%from quantum criticality,
%{\color{red}Not sure why we should emphasize ``away from criticality''}
In particular,
we examine the Hamilton-Jacobi equation.
As before, we gauge-fix and consider the normal coordinate system,
$d s^{2}(x,z) = d
z^{2} + g_{\mu\nu}(x,z) d x^{\mu} d x^{\nu}$,
i.e., $\mathcal{N}^{\mu}=0$ and $\mathcal{N}=1$.
Furthermore,
we assume that vacuum fluctuations of high-energy matter fields,
performing the gradient expansion,
%are geometrized by the gradient expansion and
are
approximated by the Einstein-Hilbert action,
$
\mathcal{H}_g = \frac{1}{2\kappa}
\sqrt{g}
\left(  R - 2 \Lambda \right)
$.
(For the explicit evaluation of $\mathcal{H}_g$
for specific models,
see \cite{Einstein_Klein_Gordon_RG_Kim,Einstein_Dirac_RG_Kim,RG_GR_Geometry_I_Kim,RG_GR_Geometry_II_Kim}.)
The holographic dual effective field theory is then given by
\begin{align}
  Z &= \int D \psi_{\alpha}(x) D g_{\mu\nu}(x,z) D \pi^{\mu\nu}(x,z) ~ \delta\Big(g^{\mu\nu}(x,0) - g^{\mu\nu}_{B}(x)\Big)
      \exp\bigg[ - \int d^{D} x \sqrt{g} ~ \mathcal{L}[\psi_{\alpha}, g^{\mu\nu}]\bigg|_{z=z_{f}}
      \nn
  &\, \qquad
      - N \int_{0}^{z_{f}} d z \int d^{D} x \Big\{ \pi^{\mu\nu} \Big( \partial_{z} g_{\mu\nu} - \beta_{\mu\nu}^{g}[g] \Big)
      + \frac{\lambda}{2} \frac{1}{\sqrt{g}} \pi^{\mu\nu} \mathcal{G}_{\mu\nu\rho\gamma} \pi^{\rho\gamma}
    + \frac{1}{2 \kappa} \sqrt{g} \Big( R - 2 \Lambda \Big) \Big\} \bigg] .
\label{gauge_fixed_Z}\end{align}

% \textcolor{blue}{
  It is straightforward to find the bulk Hamilton's equation of motion at $z<z_{f}$ for metric and its conjugate momenta:
\begin{equation}
\begin{aligned}
  \partial_{z} g_{\mu\nu} - \beta_{\mu\nu}^{g} &= -\frac{\lambda}{\sqrt{g}} \mathcal{G}_{\mu\nu\rho\sigma} \pi^{\rho\sigma}\,,
  \\
  g_{\mu\rho} g_{\nu\sigma} \partial_{z}\pi^{\rho\sigma} &= - \pi^{\rho\sigma}\frac{\delta\beta_{\rho\sigma}}{\delta g^{\mu\nu}} + \frac{\lambda}{4\sqrt{g}} g_{\mu\nu} \pi^{\kappa\lambda} \mathcal{G}_{\kappa\lambda\rho\sigma} \pi^{\rho\sigma} -\pi_{\mu\sigma}\pi_{\nu}{}^{\sigma} +\frac{1}{D-1} \pi_{\mu\nu}\pi^{\rho}{}_{\rho}  + \frac{1}{2\kappa}\Big(R_{\mu\nu} -\frac{1}{2} g_{\mu\nu}R + g_{\mu\nu}\Lambda\Big)\,.
\end{aligned}\label{bulk_eom}
\end{equation}
%
%\begin{align}
%  &
%\partial_{z} g_{\mu\nu}(x,z) - \beta_{\mu\nu}^{g}[g_{\mu\nu}(x,z)] = -
%\frac{\lambda}{\sqrt{g(x,z)}} \mathcal{G}_{\mu\nu\rho\gamma}(x,z)
%\pi^{\rho\gamma}(x,z),
%  \nonumber \\
%  &
%\partial_{z} \pi_{\mu\nu}(x,z)
%= - \frac{\lambda}{4} \frac{1}{\sqrt{g(x,z)}}
%g_{\mu\nu}(x,z) \pi^{\alpha\beta}(x,z) \mathcal{G}_{\alpha\beta\rho\gamma}(x,z)
%\pi^{\rho\gamma}(x,z) - \pi^{\alpha\beta}(x,z) \Big( \frac{\delta}{\delta
%  g^{\mu\nu}(x,z)} \beta_{\alpha\beta}^{g}[g_{\alpha\beta}(x,z)] \Big)
%\nn
%  & \qquad
%    \qquad
%    \qquad
%    + \frac{\sqrt{g(x,z)}}{2 \kappa} \Big(R_{\mu\nu}(x,z) - \frac{1}{2}
%R(x,z) g_{\mu\nu}(x,z) + \Lambda g_{\mu\nu}(x,z)\Big),
%\end{align}
Note that the boundary equations of motion at $z=z_{f}$ is given by the variation of $\delta g_{\mu\nu}(x,z_{f})$
\begin{equation}
  -\int d^{D}x\, \delta g_{\mu\nu}(x,z_{f}) \bigg[\frac{\delta}{\delta g_{\mu\nu}}\Big(\sqrt{g} \mathcal{L}[\psi_{\alpha},g^{\mu\nu}] \Big) + N \pi^{\mu\nu}\bigg]_{z=z_{f}} = 0\,,
\label{bdry_eom_derivation}
\end{equation}
where the second term arises from the variation of $g_{\mu\nu}$ in the second line of \eqref{gauge_fixed_Z}. Using the definition of the energy-momentum tensor in \eqref{Energy_Momentum_Tensor}, we get the boundary condition for $\pi^{\mu\nu}(x,z_{f})$
\begin{equation}
  \pi^{\mu\nu}(x,z_{f}) = -\frac{\sqrt{g}}{2 N}\big\langle T^{\mu\nu}(x,z_{f})\big\rangle.
\label{bdry_eom_pi}
\end{equation}
If we substitute this relation into the first equation of \eqref{bulk_eom}, we have
\begin{equation}
  \partial_{z} g_{\mu\nu}(x,z_{f}) -
\beta_{\mu\nu}^{g}(x,z_{f}) = \frac{\lambda}{2 N} \mathcal{G}_{\mu\nu\rho\gamma}(x,z_{f}) \big\langle T^{\rho\gamma}(x,z_{f})\big\rangle.
\label{IR bdry cond 1}
\end{equation}
%
%(KH: The previous Hamiltonian equations were wrong, so I corrected. I've also changed the derivation of the boundary condition to clarify that it is derived from the boundary eom. If you don't like this change, just restore the previous one. $\rightarrow$ Thanks, KH, and I (KS) like it.)}
%
%These equations are supplemented by the following boundary
%conditions
%that we obtain by varying the effective action
%with respect to
%$\pi^{\mu\nu}|_{z=z_{f}}$ and $g_{\mu\nu}|_{z=z_{f}}$:
%\begin{align}
%  \label{IR bdry cond 1}
%  &
%\partial_{z} g_{\mu\nu}(x,z_{f}) -
%\beta_{\mu\nu}^{g}[g](x,z_{f}) = \frac{\lambda}{2 N}
%\mathcal{G}_{\mu\nu\rho\gamma}(x,z_{f}) \Big\langle T^{\rho\gamma}(x,z_{f})
%\Big\rangle,
%   \\
%  \label{IR bdry cond}
%  &
%    \frac{\delta }{\delta g_{\mu\nu}(x,z_{f})} \Big\{ N \pi^{\mu\nu}(x,z_{f}) g_{\mu\nu}(x,z_{f}) + \sqrt{g(x,z_{f})} \mathcal{L}[\psi_{\alpha}(x), g^{\mu\nu}(x,z_{f})] \Big\} = 0.
%\end{align}
We point out that this holographic dual effective field theory is reduced to that of the $AdS_{D+1}$/$CFT_{D}$ duality conjecture when $\beta_{\mu\nu}^{g}[g_{\mu\nu}(x,z)] = 0$ regardless of $z$. This indicates that $\mathcal{L}[\psi_{\alpha}(x), g^{\mu\nu}(x,z_{f})]$ remains at its conformally invariant fixed point under the RG transformation, which corresponds to a special case.

This set of the Hamilton's equation of motion can
also be reformulated as an Euler-Lagrange equation of motion. Performing the path integral with respect to the canonical momentum $\pi^{\mu\nu}(x,z)$, we obtain an effective Lagrangian as follows
\begin{align}
  &
    Z = \int D \psi_{\alpha}(x) D g^{\mu\nu}(x,z) ~ \delta\Big(g^{\mu\nu}(x,0) - g^{\mu\nu}_{B}(x)\Big) ~ \exp\Big[ - \int d^{D} x \sqrt{g(x,z_{f})} ~ \mathcal{L}[\psi_{\alpha}(x), g^{\mu\nu}(x,z_{f})]
    \nonumber \\
  & \qquad
    - N \int_{0}^{z_{f}} d z \int d^{D} x \sqrt{g(x,z)}
    \Big\{ - \frac{1}{2 \lambda}
    %\Big( \partial_{z} g_{\mu\nu}(x,z) - \beta_{\mu\nu}^{g}[g_{\mu\nu}(x,z)]
    %\Big)
    V_{\mu\nu}(x,z)
    \mathcal{G}^{\mu\nu\rho\gamma}(x,z)
    V_{\rho\gamma}(x,z)
    %\Big( \partial_{z} g_{\rho\gamma}(x,z) - \beta_{\rho\gamma}^{g}[g_{\rho\gamma}(x,z)] \Big)
   + \frac{1}{2 \kappa} \Big( R(x,z) - 2 \Lambda \Big) \Big\} \Big] \,,
\end{align}
where $\mathcal{G}^{\mu\nu\rho\sigma}$ is the inverse DeWitt metric satisfying $\mathcal{G}_{\alpha\beta\gamma\delta} \mathcal{G}^{\gamma\delta\mu\nu} = \delta_{(\alpha}{}^{\mu}\delta_{\beta)}{}^{\nu}$
\begin{equation}
  \mathcal{G}^{\mu\nu\rho\sigma}= \frac{1}{2} g^{\mu\rho} g^{\nu\sigma} + \frac{1}{2} g^{\nu\rho} g^{\mu\sigma} - g^{\mu\nu} g^{\rho\sigma}\,,
\label{}\end{equation}
and we introduced an auxiliary field $V_{\mu\nu}$,
\begin{equation}
  V_{\mu\nu}(x,z) = \partial_{z}g_{\mu\nu}(x,z) - \beta^{g}_{\mu\nu}[g_{\mu\nu}(x,z)]\,,
  \label{}
\end{equation}
to lighten notations.

Accordingly, the Euler-Lagrange equation is given by
%\bqa && - \frac{1}{\sqrt{g(x,z)}} ~ \partial_{z} \Bigg( \frac{\sqrt{g(x,z)}}{\lambda} \mathcal{G}_{\mu\nu}^{\rho\gamma}(x,z) \Big( \partial_{z} g_{\rho\gamma}(x,z) - \beta_{\rho\gamma}^{g}[g_{\rho\gamma}(x,z)] \Big) \Bigg) \nn && = - \frac{g_{\mu\nu}(x,z)}{4\lambda} \Big( \partial_{z} g_{\alpha\beta}(x,z) - \beta_{\alpha\beta}^{g}[g_{\alpha\beta}(x,z)] \Big) \mathcal{G}^{\alpha\beta\rho\gamma}(x,z) \Big( \partial_{z} g_{\rho\gamma}(x,z) - \beta_{\rho\gamma}^{g}[g_{\rho\gamma}(x,z)] \Big) \nn && + \frac{1}{\lambda} \eta_{\mu\nu\alpha\beta}(x,z) \mathcal{G}^{\alpha\beta\rho\gamma}(x,z) \Big( \partial_{z} g_{\rho\gamma}(x,z) - \beta_{\rho\gamma}^{g}[g_{\rho\gamma}(x,z)] \Big) + \frac{1}{2 \kappa} \Big(R_{\mu\nu}(x,z) - \frac{1}{2} R(x,z) g_{\mu\nu}(x,z) + \Lambda g_{\mu\nu}(x,z)\Big) , \label{Bulk_Eq_Metric} \eqa
%
\begin{equation}
\begin{aligned}
  & \frac{g_{\mu\alpha} g_{\nu\beta} }{\sqrt{g}} \partial_{z} \bigg( \frac{\sqrt{g}}{\lambda} \mathcal{G}^{\alpha\beta\gamma\delta} V_{\gamma\delta} \bigg)
  - \bigg(\frac{g_{\mu\nu} V_{\alpha\beta}}{4\lambda} - \frac{\eta_{\mu\nu\alpha\beta}}{\lambda} \bigg)\mathcal{G}^{\alpha\beta\gamma\delta} V_{\gamma\delta}
  \\
  &
  \qquad
  + \frac{ g^{\rho\sigma}}{\lambda} \Big(V_{\mu\rho} V_{\nu\sigma} -V_{\mu\nu} V_{\rho\sigma}\Big)
   + \frac{1}{2 \kappa} \Big(R_{\mu\nu} - \frac{1}{2}  g_{\mu\nu} R+ \Lambda g_{\mu\nu}\Big)=0\,. \label{Bulk_Eq_Metric}
\end{aligned}
\end{equation}
Here,
we introduce
%
%the viscosity tensor,
%
%{\color{red}(Can put motivation to call this viscosity tensor?)}
%
\begin{align}
  \label{viscosity}
\eta_{\mu\nu\alpha\beta}(x,z) \equiv \frac{\delta}{\delta
  g^{\mu\nu}(x,z)} \beta_{\alpha\beta}^{g}[g_{\alpha\beta}(x,z)] .
\end{align}
We recall that the RG $\beta$-function for the metric tensor
is given by the energy-momentum tensor of the renormalized QFT
at a given RG scale $z$.
It is natural to call $\eta_{\mu\nu\alpha\beta}(x,z)$
the viscosity tensor,
as it is given by the derivative of the energy-momentum tensor
with respect to the metric tensor.
The role of this viscosity tensor in the dynamics of metric fluctuations will be discussed below.

We now discuss the self-consistency of the present
formulation, in particular, the consistency
of the RG invariance, and the Hamilton (Hamilton-Jacobi) equations
%Self-consistency of the present formulation can be seen in the Hamilton-Jacobi equation \cite{Holographic_RG_Review}.
\cite{Holographic_RG_Review}.
The generating functional has to be invariant under the RG transformation
in the sense that
\begin{align}
  \frac{d}{d z_{f}} \ln Z = 0 .
\end{align}
This holds true if the following condition is satisfied
\begin{align}
  \label{RG inv}
  & \Big( \partial_{z_{f}} g_{\mu\nu}(x,z_{f}) \Big) \frac{1}{N} \frac{\partial }{\partial g_{\mu\nu}(x,z_{f})} \Big\langle \sqrt{g(x,z_{f})} \mathcal{L}[\psi_{\alpha}(x), g^{\mu\nu}(x,z_{f})] \Big\rangle + \frac{1}{N} \partial_{z_{f}} \Big\langle \sqrt{g(x,z_{f})} \mathcal{L}[\psi_{\alpha}(x), g^{\mu\nu}(x,z_{f})] \Big\rangle
    \nn
  & + \pi^{\mu\nu}(x,z_{f}) \Big( \partial_{z_{f}} g_{\mu\nu}(x,z_{f}) - \beta_{\mu\nu}^{g}[g_{\mu\nu}(x,z_{f})] \Big) + \frac{\lambda}{2} \frac{1}{\sqrt{g(x,z_{f})}} \pi^{\mu\nu}(x,z_{f}) \mathcal{G}_{\mu\nu\rho\gamma}(x,z) \pi^{\rho\gamma}(x,z_{f})
    \nn
  & + \frac{1}{2 \kappa} \sqrt{g(x,z_{f})} \Big( R(x,z_{f}) - 2 \Lambda \Big) = 0 .
\end{align}
One may regard this equation as the Callan-Symanzik equation \cite{QFT_Textbook}
for the free-energy functional in the large $N$ limit.
We recall the IR boundary canonical momentum tensor is given by Eq.\ \eqref{bdry_eom_derivation}.
%
%\textcolor{blue}{
%KH:We may remove the following equation because it is already mentioned in \eqref{bdry_eom_pi}:
%\\
%We note that, from the IR boundary condition \eqref{IR bdry cond},
%we obtain the IR boundary canonical momentum tensor as
%\bqa && \pi^{\mu\nu}(x,z_{f}) = - \frac{1}{N} \frac{\delta }{\delta g_{\mu\nu}(x,z_{f})} \Big\langle \sqrt{g(x,z_{f})} \mathcal{L}[\psi_{\alpha}(x), g^{\mu\nu}(x,z_{f})] \Big\rangle = - \frac{\sqrt{g(x,z_{f})}}{2 N} \Big\langle T^{\mu\nu}(x,z_{f}) \Big\rangle \label{IR_BC} \eqa
%in the large $N$ limit.
%combined with the Hamilton's equation of motion for the metric tensor}
%
In addition, it is natural to assume $\partial_{z_{f}} \Big\langle
\sqrt{g(x,z_{f})} \mathcal{L}[\psi_{\alpha}(x), g^{\mu\nu}(x,z_{f})] \Big\rangle
= 0$, i.e., the IR boundary Lagrangian does not depend
explicitly on the boundary coordinate $z_{f}$,
since
all the cutoff dependence
is through the running of the coupling constants as a function of $z$.
%
%{\color{blue}
%(By the way, it is not clear to me why $\partial_{z_{f}} \Big\langle
%\sqrt{g(x,z_{f})} \mathcal{L}[\psi_{\alpha}(x), g^{\mu\nu}(x,z_{f})] \Big\rangle
%= 0$. Could you explain more details?)}
%{\color{red}
%  (SR: I thought it is because all the cutoff dependence
%  are through the running of the coupling constants as a function of
%  $z$?)
%}
%
As a result, the RG invariance of the free energy
  \eqref{RG inv}
is reduced to the Hamilton-Jacobi equation
%{\color{red}(Why the following equation is called the HJ equation?)}
\begin{align}
0 = \frac{\lambda}{2} \frac{1}{\sqrt{g(x,z_{f})}} \pi^{\mu\nu}(x,z_{f})
\mathcal{G}_{\mu\nu\rho\gamma}(x,z) \pi^{\rho\gamma}(x,z_{f}) -
\pi^{\mu\nu}(x,z_{f}) \beta_{\mu\nu}^{g}[g_{\mu\nu}(x,z_{f})] + \frac{1}{2
  \kappa} \sqrt{g(x,z_{f})} \Big( R(x,z_{f}) - 2 \Lambda \Big),
 \label{Hamilton_Jacobi_Eq}
\end{align}
where $\pi^{\mu\nu}= \delta S/\delta g_{\mu\nu}$.
%{\color{red}
%  (To make this look more like the Hamilton-Jacobi, do we need to set $\pi^{\mu\nu}= \delta S/\delta g_{\mu\nu}$
%  and rewrite the equation
%  in terms of $S$? $\rightarrow$ Basically, I agree with this statement. One minor concern is the expression is dirty. Actually, Eq. (24) (up to the first equality) is what you said. let's consider more on this replacement.)}
%  {\color{red}(Also, is it ok to have $z_f$ here instead of generic $z$?
%    $\rightarrow$ I also agree with this statement since $z_{f}$ can be
%    arbitrary. But, inside $z_{f}$, i.e., $z < z_{f}$, I do not think that this
%    Hamilton-Jacobi equation holds. Is this statement correct,
%    Kanghoon?}\textcolor{blue}{$\to$ I believe we cannot replace $z_{f}$ to $z$,
%    because it does not have the same meaning and we cannot use the boundary
%    condition anymore.
%    %By the way, it is not clear to me why $\partial_{z_{f}} \Big\langle
%    %\sqrt{g(x,z_{f})} \mathcal{L}[\psi_{\alpha}(x), g^{\mu\nu}(x,z_{f})]
%    %\Big\rangle = 0$. Could you explain more details?
%  })
We emphasize that the solution of this Hamilton-Jacobi equation is given by the
IR boundary condition [Eq. \eqref{bdry_eom_derivation}], where $\mathcal{L}[\psi_{\alpha}(x),
g^{\mu\nu}(x,z_{f})]$ is the IR boundary effective Lagrangian determined
self-consistently. Again, this Hamilton-Jacobi equation becomes that of the
${\it AdS}_{D+1}/{\it CFT}_{D}$ duality conjecture when $\beta_{\mu\nu}^{g}[g_{\mu\nu}(x,z)] = 0$.

As a further consistency check,
% Before going to the section of discussions,
we point out that the holographic
dual effective field theory will follow
the constraint.
%
%the Ward identity. {\color{red}(the constraint? $\rightarrow$ Actually, the term of the Ward identity is not mine. I think that this Ward identity is well summarized in Ref. \cite{Holographic_Ward_Identity}.)} 
% 
Inserting the Hamilton's equation for the metric tensor
into the constraint 
%into the Ward identity,
$\mathcal{D}_{\nu} \pi^{\mu\nu}(x,z) = 0$,
it is natural to expect that the covariant derivative for the metric tensor would vanish \cite{Holographic_Ward_Identity}, regarded to be a part of full equations of motion \cite{GR_Textbook},
%
%nothing but the metric compatibility relation \cite{GR_Textbook},
%
and that for the RG
$\beta$-function also becomes zero, nothing but the energy-momentum
tensor-current conservation law. It seems to be a natural generalization to introduce the RG $\beta$-function of the metric tensor into the bulk effective action for gravity, expected to work away from quantum criticality.
% 
%(\textcolor{blue}{KH: In fact, the constraint $\mathcal{D}_{\nu} \pi^{\mu\nu}(x,z) = 0$ is not satisfied automatically for an arbitrary metric tensor, because this is a part of the full equations of motion, which is $R_{0i}$ components. Thus we have to consider a specific solution of the Einstein equation to verify it. On the other hand, the metric compatibility cannot be applied here, since $\partial_{z}$ and $D_{\mu}$ are not commute to each other. Obviously it gives a nontrivial check for the validity, but it is not satisfied ``automatically".})
%

\subsection{Discussion}

\subsubsection{Entanglement entropy perspectives}

The present RG-reformulated dual gravity action may be reinterpreted in perspectives of entanglement entropy \cite{Entanglement_Entropy,Entanglement_Entropy_Calabrese_Cardy_I,Entanglement_Entropy_Calabrese_Cardy_II,Entanglement_Entropy_Review_III,
Entanglement_Entropy_Review_IV}. The entanglement entropy is given by
\bqa && S_{EE}^{UV} = S_{EE}^{GR}(z_{f}) + S_{EE}^{M}(z_{f}) \eqa
in our holographic dual effective field theory. Here, $S_{EE}^{UV}$ is the
entanglement entropy of the UV effective QFT defined at the UV boundary $z = 0$.
$S_{EE}^{GR}(z_{f})$ is the entanglement entropy of the emergent dual gravity
bulk action with the IR boundary $z = z_{f}$. $S_{EE}^{M}(z_{f})$ is that of the
IR boundary action at $z = z_{f}$.
This seemingly natural formula can be derived
from the holographic dual effective field theory
\cite{Einstein_Klein_Gordon_RG_Kim}
using the replica trick \cite{Entanglement_Entropy_Heat_Kernel,Generalized_Gravitational_Entropy}.
%Although this formula looks natural, considering the holographic dual effective field theory,y it can be derived \cite{Einstein_Klein_Gordon_RG_Kim} based on the replica trick \cite{Entanglement_Entropy_Heat_Kernel,Generalized_Gravitational_Entropy}.

Since the entanglement entropy of the UV effective QFT does not depend on the IR boundary coordinate $z_{f}$, we obtain the following Callan-Symanzik equation for the entanglement entropy as
\bqa && 0 = \partial_{z_{f}} S_{EE}^{GR}(z_{f}) + \partial_{z_{f}} S_{EE}^{M}(z_{f}) . \eqa

Resorting to the replica trick for the gravitational effective action \cite{Generalized_Gravitational_Entropy}, one may argue that $S_{EE}^{GR}(z_{f})$ is given by an area of the Ryu-Takayanagi minimal surface \cite{Entanglement_Entropy_Ryu_Takayanagi_I,Entanglement_Entropy_Ryu_Takayanagi_II}. Since we did not address the role of the RG $\beta-$function in the Ryu-Takayanagi minimal surface yet, we used the term of ``may argue". Essentially the same replica trick gives rise to the area law of $S_{EE}^{M}(z_{f})$ \cite{Entanglement_Entropy_Heat_Kernel} at the IR boundary. Here, we represent both entanglement entropies as follows
\bqa && S_{EE}^{GR}(z_{f}) = \frac{\mathcal{A}_{RT}(z_{f})}{4 G_{D+1}} , ~~~~~ S_{EE}^{M}(z_{f}) = \frac{\mathcal{A}_{QFT}(z_{f})}{4 G_{D}} . \eqa
$\mathcal{A}_{RT}(z_{f})$ is a $(D-1)$-dimensional Ryu-Takayanagi minimal-surface area at $z = z_{f}$, and $G_{D+1}$ is $(D+1)$-dimensional Newton constant in $S_{EE}^{GR}(z_{f})$. $\mathcal{A}_{QFT}(z_{f})$ is a $(D-2)$-dimensional surface area of the QFT with renormalization at $z = z_{f}$, and $G_{D}$ is $D$-dimensional Newton constant in $S_{EE}^{M}(z_{f})$.

As a result, we obtain
\bqa && 0 = \frac{\partial_{z_{f}} \mathcal{A}_{RT}(z_{f})}{4 G_{D+1}} + \frac{\partial_{z_{f}} \mathcal{A}_{QFT}(z_{f})}{4 G_{D}} . \eqa
This area formulation interprets the appearance of the RG-reformulated dual
gravity action in a geometrical way.
The decrease of the $(D-2)$-dimensional surface area of the QFT,
representing the decrease of the entanglement entropy in the QFT,
gives rise to the increase of the $(D-1)$-dimensional Ryu-Takayanagi
minimal-surface area,
describing the increase of the entanglement entropy in the bulk gravity theory,
where the RG transformation is performed at $z = z_{f}$. It would be interesting to show this relation explicitly.

\subsubsection{Role of the viscosity tensor in the dynamics of metric fluctuations}

Finally, we discuss the role of the RG $\beta$-function in the bulk dynamics of
metric fluctuations. It is not easy to solve Eq.\ (\ref{Bulk_Eq_Metric})
and find the RG flow of the metric tensor because the RG $\beta$-function gives rise to higher-curvature corrections in the $D$-dimensional Einstein-Hilbert action. Performing the gradient expansion in Eq.\ (\ref{beta_ft}) with Eq.\ (\ref{Energy_Momentum_Tensor}) \cite{Entanglement_Entropy_Heat_Kernel}, one can express the average of the energy-momentum tensor in terms of curvature tensors, which results in higher-curvature terms in the case of $\lambda \not= 0$. Here, we consider a near fixed-point solution of the metric tensor, which allows us to investigate the bulk dynamics of metric fluctuations in a linearized fashion around the fixed-point background geometry.

We recall the IR boundary condition \eqref{IR bdry cond 1}.
Taking the $z_{f} \rightarrow \infty$ limit, quantum fluctuations of matter
fields are integrated out completely. As a result, the average of the
energy-momentum tensor cannot but vanish.
Since the RG $\beta$-function also vanishes, the resulting IR boundary condition is given by
\bqa && \lim_{z_{f} \rightarrow \infty} \partial_{z} g_{\mu\nu}(x,z) \Big|_{z = z_{f}} = 0 . \eqa
%
%\bqa && \lim_{z_{f} \rightarrow \infty} \partial_{z} g_{\mu\nu}(x,z) \Big|_{z = z_{f}} = 0 , ~~~~~ \lim_{z_{f} \rightarrow \infty} \beta_{\mu\nu}^{g}[g_{\mu\nu}(x,z_{f})] = 0 \eqa
%
In this limit, the RG flow equation of the metric tensor is reduced to
%
%\bqa && - \mathcal{G}_{\mu\nu\rho\gamma}(x,z) \partial_{z}^{2} g^{\rho\gamma}(x,z) = \frac{\lambda}{2 \kappa} \Big(R_{\mu\nu}(x,z) - \frac{1}{2} R(x,z) g_{\mu\nu}(x,z) + \Lambda g_{\mu\nu}(x,z)\Big) . \eqa
%
\begin{equation}
  \partial_{z} V_{\mu\nu}(x,z) - g_{\mu\nu}(x,z) g^{\rho\sigma}(x,z) \partial_{z} V_{\rho\sigma}(x,z) = \frac{\lambda}{2 \kappa \sqrt{g(x,z)}} \Big(R_{\mu\nu}(x,z) - \frac{1}{2} R(x,z) g_{\mu\nu}(x,z) + \Lambda g_{\mu\nu}(x,z) \Big) .
\label{}
\end{equation}
We recall $V_{\mu\nu}(x,z) = \partial_{z} g_{\mu\nu}(x,z) - \beta^{g}_{\mu\nu}[g_{\mu\nu}(x,z)]$.
%{\color{red}(How do we get this equation? $\rightarrow$ Inserting both conditions of Eq. (30) and vanishing beta function in the IR limit into Eq. (14) (the equation of motion for the metric tensor), I obtained this result. This equation has to be the same as that of the conventional AdS/CFT duality since the beta function vanishes here. Kanghoon, please confirm this equation.} \textcolor{blue}{KH: From the boundary condition \eqref{IR bdry cond 1}, $\pi^{\mu\nu}$ also vanishes. In this case, the only remained part in the RHS of the $\partial_{z} \pi$ equation is the Einstein equation, $\partial_{z} \pi_{\mu\nu} \sim R_{\mu\nu} -\frac{1}{2} g_{\mu\nu} R + g_{\mu\nu}\Lambda$. However, we have to be careful with the index position and the equation should be
%
%\begin{equation}
%  \partial_{z}^{2}g_{\mu\nu} - \partial_{z}\beta^{g}_{\mu\nu} -g_{\mu\nu} g^{\rho\sigma} \big(\partial_{z}^{2}g_{\rho\sigma}-\partial_{z}\beta^{g}_{\rho\sigma}\big)= \frac{\lambda}{2 \kappa \sqrt{g}} \Big(R_{\mu\nu} - \frac{1}{2} R g_{\mu\nu} + \Lambda g_{\mu\nu}\Big) \nonumber
%\label{}
%\end{equation}
%%
%Here, I kept $\partial_{z}\beta^{g}_{\mu\nu}$ term just in case. Do you think is it always vanish?)}
This equation is identical to that of the conventional holography if $\lim_{z_{f} \rightarrow \infty} \partial_{z} \beta^{g}_{\mu\nu}[g_{\mu\nu}(x,z)] \Big|_{z = z_{f}} = 0$ is assumed near an IR fixed point. Here, gauge fixing is assumed as discussed before.
%
%\bqa && g_{\mu\nu}(x,z) = g_{\mu\nu}^{BH}(x,z) \eqa
%
The background solution is an (thermal)
${\it AdS}_{D+1}$ geometry at zero temperature (below the Hawking-Page transition
  temperature) and an ${\it AdS}_{D+1}$ black hole above the Hawking-Page transition temperature \cite{Hawking_Page_Transition}.

Considering small fluctuations around this background geometry $\tilde{g}_{\mu\nu}$ as
%\bqa && g_{\mu\nu}(x,z) = g_{\mu\nu}^{BH}(x,z) + h_{\mu\nu}(x,z) , \eqa
%
\begin{equation}
  g_{\mu\nu}(x,z) = \tilde{g}_{\mu\nu}(x,z) + h_{\mu\nu}(x,z) \,,
\label{}\end{equation}
the linearized ``Einstein" equation for the metric tensor is given by
%\bqa && - \mathcal{G}_{\mu\nu\rho\gamma}(x,z) \partial_{z}^{2} h^{\rho\gamma}(x,z) = - \eta_{\mu\nu\alpha\beta}^{BH}(x,z) \mathcal{G}^{\alpha\beta\rho\gamma}(x,z) \eta_{\rho\gamma\sigma\delta}^{BH}(x,z) h^{\sigma\delta}(x,z) \nn && + \frac{\lambda}{2 \kappa} \frac{g_{\alpha\beta}^{BH}(x,z) h^{\alpha\beta}(x,z)}{[\sqrt{g_{BH}(x,z)}]^{2}} \Big(R_{\mu\nu}^{BH}(x,z) - \frac{1}{2} R_{BH}(x,z) g_{\mu\nu}^{BH}(x,z) + \Lambda g_{\mu\nu}^{BH}(x,z)\Big) \nn && + \frac{\lambda}{4 \kappa} \Big(\partial_{\sigma} \partial_{\mu} h_{\nu}^{\sigma}(x,z) + \partial_{\sigma} \partial_{\nu} h_{\mu}^{\sigma}(x,z) - \partial_{\mu} \partial_{\nu} [g_{\rho\lambda}^{BH}(x,z) h^{\rho\lambda}(x,z)] - g^{\alpha\beta}_{BH}(x,z) \partial_{\alpha} \partial_{\beta} h_{\mu\nu}(x,z) \nn && - g_{\mu\nu}^{BH}(x,z) \partial_{\rho} \partial_{\lambda} h^{\rho\lambda}(x,z) + g_{\mu\nu}^{BH}(x,z) g^{\alpha\beta}_{BH}(x,z) \partial_{\alpha} \partial_{\beta} [g_{\rho\lambda}^{BH}(x,z) h^{\rho\lambda}(x,z)] + 2 \Lambda h_{\mu\nu}(x,z)\Big) . \eqa
%
\begin{equation}
\begin{aligned}
   & \frac{\tilde{g}_{\mu\alpha}\tilde{g}_{\nu\beta}}{2\sqrt{\tilde{g}}} \partial_{z} \bigg[\frac{\sqrt{\tilde{g}}}{\lambda} \Big(\tilde{g}^{\rho\sigma}h_{\rho\sigma} \tilde{\mathcal{G}}^{\alpha\beta\gamma\delta}\tilde{V}_{\gamma\delta}
  +2\big(\delta_{h}\tilde{\mathcal{G}}^{\alpha\beta\gamma\delta} \big) \tilde{V}_{\gamma\delta}
  +2\tilde{\mathcal{G}}^{\alpha\beta\gamma\delta} \big(\delta_{h}\tilde{V}_{\gamma\delta}\big)\Big) \bigg]
  \\
  &+\bigg(\frac{2h_{(\mu|\alpha|}\tilde{g}_{\nu)\beta}}{\sqrt{\tilde{g}}} - \frac{\tilde{g}_{\mu\alpha}\tilde{g}_{\nu\beta}}{2\sqrt{\tilde{g}}}\bigg) \partial_{z}\bigg[\frac{\sqrt{\tilde{g}}}{\lambda} \tilde{\mathcal{G}}^{\alpha\beta\gamma\delta} \tilde{V}_{\gamma\delta}\bigg]
  - \bigg(\frac{\tilde{g}_{\mu\nu}\tilde{V}_{\alpha\beta}}{2\lambda} + \frac{\tilde{\eta}_{\mu\nu\alpha\beta}}{\lambda}\bigg)\tilde{V}_{\alpha\beta} \tilde{\mathcal{G}}^{\alpha\beta\gamma\delta} \Big(\partial_{z}h_{\gamma\delta}+h^{\rho\sigma} \tilde{\eta}_{\rho\sigma\gamma\delta}\Big)
  \\
  &-\bigg(\frac{\tilde{g}_{\mu\nu}}{4\lambda} \tilde{V}_{\alpha\beta}+\frac{\tilde{\eta}_{\mu\nu\alpha\beta}}{\lambda}\bigg)\big(\delta_{h}\tilde{\mathcal{G}}^{\alpha\beta\gamma\delta} \big) \tilde{V}_{\gamma\delta}
   -\frac{h_{\mu\nu}}{4\lambda} \tilde{V}_{\alpha\beta}\tilde{\mathcal{G}}^{\alpha\beta\gamma\delta} \tilde{V}_{\gamma\delta}
   -\frac{h^{\rho\sigma}}{\lambda}\Big(\tilde{V}_{\mu\rho}\tilde{V}_{\nu\sigma}-\tilde{V}_{\mu\nu}\tilde{V}_{\rho\sigma}\Big)
   \\
   &+\frac{\tilde{g}^{\rho\sigma}}{\lambda} \Big(
    \tilde{V}_{\mu\rho}\big(\delta_{h}\tilde{V}_{\nu\rho}\big)
   +\big(\delta_{h}\tilde{V}_{\mu\rho}\big)\tilde{V}_{\nu\rho}
   -\tilde{V}_{\mu\nu}\big(\delta_{h}\tilde{V}_{\rho\sigma}\big)
   -\big(\delta_{h}\tilde{V}_{\mu\nu}\big)\tilde{V}_{\rho\sigma}\Big)
   \\
   &+\frac{1}{4\kappa} \bigg[\tilde{\nabla}^{\rho} \Big(\tilde{\nabla}_{\mu} h_{\rho\nu} {+}\tilde{\nabla}_{\nu} h_{\rho\mu} {-}\tilde{\nabla}_{\rho} h_{\mu\nu}\Big)-\tilde{\nabla}_{\nu} \tilde{\nabla}_{\mu} h^{\rho}{}_{\rho}
   +\tilde{g}_{\mu\nu}\Big(h^{\rho\sigma} \tilde{R}_{\rho\sigma} {-}\tilde{\nabla}_{\rho} \tilde{\nabla}_{\sigma} h^{\rho\sigma} {+}\tilde{\nabla}^{\rho} \tilde{\nabla}_{\rho} h^{\sigma}{}_{\sigma}\Big) -h_{\mu\nu}\big(\tilde{R}-2\Lambda\big)\bigg]=0
\end{aligned}\label{}
\end{equation}
where $\tilde{\mathcal{G}}^{\alpha\beta\gamma\delta}$ and $\tilde{V}_{\mu\nu}$ are the background quantities and
\begin{equation}
  \delta_{h}\tilde{\mathcal{G}}^{\alpha\beta\gamma\delta} = - \big(h^{\alpha(\gamma}\tilde{g}^{\delta)\beta} + h^{\beta(\gamma}\tilde{g}^{\delta)\alpha} -h^{\alpha\beta}\tilde{g}^{\gamma\delta} -\tilde{g}^{\alpha\beta}h^{\gamma\delta} \big) \,,
 \qquad
 \delta_{h} \tilde{V}_{\alpha\beta} = \partial_{z}h_{\alpha\beta}+h^{\rho\sigma} \tilde{\eta}_{\rho\sigma\alpha\beta}\,.
\label{}\end{equation}
Here,
$\tilde{\eta}_{\mu\nu\alpha\beta}(x,z)$
is the viscosity tensor
with a background black hole geometry,
$\tilde{\eta}_{\mu\nu\alpha\beta}(x,z) \equiv
\frac{\delta}{\delta g^{\mu\nu}(x,z)}
\beta_{\alpha\beta}^{g}[g_{\alpha\beta}(x,z)] \Big|_{g^{\mu\nu}(x,z) =
  \tilde{g}^{\mu\nu}(x,z)}$
(see \eqref{viscosity}).
This may not vanish near the fixed point $z_{f} \rightarrow \infty$
while the RG $\beta$-function itself becomes zero.
It is interesting to observe
that this viscosity tensor can result in instability of metric fluctuations near
the fixed-point background geometry.
We speculate that this potential instability originates from higher curvature
corrections to the Einstein-Hilbert action \cite{Higher_Curvature_GR}. More
generally, we suspect that the RG $\beta$-function of the metric tensor may
encode the so-called Ricci flow
\cite{Ricci_Flow_0,Ricci_Flow_I,Ricci_Flow_II,Ricci_Flow_III,Ricci_Flow_IV,Ricci_Flow_V}.
The Ricci flow equation is to describe the deformation of a Riemannian metric
$g_{\mu\nu}(x,z)$ with an extra-dimensional space coordinate $z$,
which plays the same role as time.
This evolution equation may be regarded as an analog
of the diffusion equation for geometries, given by a parabolic partial
differential equation. The deformation is governed by
the Ricci curvature, and
leads to homogeneity of geometry.
In principle, one may consider that this Ricci flow equation
arises from the gradient expansion of the Green's function with respect to the
mass parameter \cite{Einstein_Klein_Gordon_RG_Kim}.
%{\color{red} (Why?? $\rightarrow$ In my opinion, this is quite an essential point. Something new here can appear from the RG beta function of the metric tensor. If we calculate the energy-momentum tensor of the renormalized QFT at $z$, and perform the gradient expansion with respect to the metric tensor in the denominator of the Green's function of the matter field, the energy-momentum tensor has to be expanded in terms of various combinations of curvature tensors. If I performed the gradient expansion correctly, diffeomorphism invariance in $D$ dimensions has to be satisfied. In other words, I am saying $(\partial_{z} g - \beta)^{2} \sim (\partial_{z} g - \mathcal{F}(R, R_{\mu\nu}, ...))^{2}$. Expanding the square, I will have curvature squares and etc., regarded to be corrections to the Einstein-Hilbert action. As far as I know, there are instabilities due to higher derivative corrections.)}
Actually, this instability of the background geometry may be interpreted as a run-away RG flow toward a fixed point different from the present one. It would be interesting to study how the universal lower bound of the ratio between the shear viscosity and the entropy \cite{Holographic_Liquid_Son_IV} is modified by this viscosity tensor
\cite{Viscosity_bound_violation}.

\section{Conclusion}

We proposed a prescription for an emergent dual holographic description of a
quantum field theory, expected to work even away from quantum criticality.
Although
% the UV-IR mapping is not exact but approximate,
we invoke the bulk locality assumption and do not include higher spin fields,
the holographic dual effective field theory takes into account quantum
corrections in a non-perturbative way
through a non-perturbative RG flow in the emergent extra-dimensional space. Self-consistency of this non-perturbative framework was claimed based on the Hamilton-Jacobi equation, the solution of which is given by the IR boundary effective action.
%
%In addition, the holographic dual effective field theory satisfies the Ward identity naturally.
%

Before closing, we point out that it is straightforward to generalize the present dual holographic description to the case with additional effective interactions. For example, one may consider either spontaneous chiral symmetry breaking or effective interactions between U(1) conserved currents. Such interactions are responsible for appearance of dual scalar fields and U(1) gauge fields, respectively, in the corresponding holographic dual effective field theory.

\begin{acknowledgments}
K.-S. Kim was supported by the Ministry of Education, Science, and Technology
(NRF-2021R1A2C1006453 and NRF-2021R1A4A3029839) of the National Research
Foundation of Korea (NRF) and by TJ Park Science Fellowship of the POSCO TJ Park
Foundation.
S.R.~is supported by the National Science Foundation under
Award No.\ DMR-2001181, and by a Simons Investigator Grant from
the Simons Foundation (Award No.~566116).
K.\ L.\ is supported by an appointment to the JRG Program at the APCTP through the Science and Technology Promotion Fund
and Lottery Fund of the Korean Government.
It is also supported by the National Research Foundation of Korea(NRF) grant
funded
by the Korea government(MSIT) No.\ 2021R1F1A1060947 and the Korean Local Governments
of Gyeongsangbuk-do Province and Pohang City.
\end{acknowledgments}


\begin{thebibliography}{9}
\bibitem{Comment_Higher_Spin_GT} One may suggest a higher-spin gauge theory \cite{Higher_Spin_Gauge_Theory_I,Higher_Spin_Gauge_Theory_II,Higher_Spin_Gauge_Theory_III,Higher_Spin_Gauge_Theory_IV,Higher_Spin_Gauge_Theory_V} as an effective holographic dual field theory. However, it turns out that the nonlocal diffeomorphism invariance, responsible for the emergence of the higher-spin gauge theory \cite{RG_Holography_VII,RG_Holography_VII_Q,RG_Holography_VIII}, is explicitly broken in general cases due to the existence of effective interactions, although the nonlocal diffeomorphism invariance exists in the absence of such interactions. See Ref. \cite{Einstein_Klein_Gordon_RG_Kim} for more discussions. In this case, the ultimate fate of higher spin fields has not been clarified as far as we know.
\end{thebibliography}
\end{document}